\documentclass[conference]{IEEEtran}
\usepackage{tikz}
\usepackage{graphicx}
\usepackage{cite}
\usepackage{amsmath, bm}
\usepackage{amsthm}
\usepackage{amssymb, epsfig}
\usepackage{subfigure,algorithmic,float}
\usepackage{lipsum}
\usepackage{textcomp}
\usepackage{epstopdf}

\newcommand{\Hnull}{\mathcal{H}_0}
\newcommand{\Halt}{\mathcal{H}_1}

\newcommand{\Honull}{\mathcal{{D}}_0}
\newcommand{\Hoalt}{\mathcal{{D}}_1}

\usepackage{cite,graphicx,amsmath,amssymb,cite,algorithm}

\begin{document}
\newtheorem{theorem}{\textbf{Theorem}}
\newtheorem{proposition}{\textbf{Proposition}}
\newtheorem{corollary}{\textbf{Corollary}}
\newtheorem{remark}{\textbf{Remark}}
\newtheorem{lemma}{\textbf{Lemma}}

\title{Covert Communications with A Full-Duplex Receiver over Wireless Fading Channels}
\author{
\IEEEauthorblockN{Jinsong~Hu$^{\ast \dag}$, Khurram Shahzad$^{\dag}$, Shihao~Yan$^{\dag}$, Xiangyun~Zhou$^{\dag}$, Feng~Shu$^{\ast}$, Jun Li$^{\ast}$}
\IEEEauthorblockA{$^{\ast}$School of Electronic and Optical Engineering, Nanjing University of Science and Technology, Nanjing, Jiangsu, China}
\IEEEauthorblockA{$^{\dag}$Research School of Engineering, The Australian National University, Canberra, ACT, Australia}
\IEEEauthorblockA{Emails:\{jinsong\_hu, shufeng, jun.li\}@njust.edu.cn,~\{khurram.shahzad, shihao.yan, xiangyun.zhou\}@anu.edu.au}
}

\maketitle
\begin{abstract}
In this work, we propose a covert communication scheme where the transmitter  attempts to hide its transmission to a full-duplex receiver, from a warden that is to detect this covert transmission using a radiometer. Specifically, we first derive the detection error rate at the warden, based on which the optimal detection threshold for its radiometer is analytically determined and its expected detection error rate over wireless fading channels is achieved in a closed-form expression. Our analysis indicates that the artificial noise deliberately produced by the receiver with a random transmit power, although causes self-interference, offers the capability of achieving a positive effective covert rate for any transmit power (can be infinity) subject to any given covertness requirement on the expected detection error rate. This work is the first study on the use of the full-duplex receiver with controlled artificial noise for achieving covert communications and invites further investigation in this regard.
\end{abstract}

\section{Introduction}
The Internet of Things (IoT) offers promising solutions to the materialization of concepts envisioning \textit{everything smart}~\cite{IoT_Comm_Mag_15}, and with this vision coming to reality in the recent years, the dependency of users on wireless devices is also rapidly increasing. Due to the broadcast nature of wireless channels, the security and privacy of wireless communications has been an ever-increasing concern, which now is the biggest barrier to the wide-spread adoption of IoT technologies~\cite{IoT_Security_2016}. Traditional security techniques offer protection against eavesdropping through encryption, guaranteeing the integrity of messages over the air. However, it has been shown in the recent years that even the most robust encryption techniques can be defeated by a determined adversary. Physical-layer security, on the other hand, exploits the dynamic characteristics of the wireless medium to minimize the information obtained by eavesdroppers \cite{bloch2011physical}. However, it does not provide protection against the detection of a transmission in the first place, which can offer an even stronger level of security, as the transmission of encrypted transmission can spark suspicion in the first place and invite further probing by skeptical eavesdroppers.

On the contrary, apart from protecting the content of communication, covert communication (also termed low probability of detection communication) aims to enable a wireless transmission between two users while guaranteeing a negligible detection probability of this transmission at a warden. Such communications are also highly desirable by government and military organizations, who are interested in keeping their activities hidden over the air. Covert communication has recently drawn significant research interests and is emerging as a cutting-edge technique in the context of wireless communication security \cite{bash2015hiding,bloch2016covert}. The fundamental limits of covert communication have been studied under various channels, such as additive white Gaussian noise (AWGN) channel \cite{bash2013limits}, binary symmetric channel (BSC) \cite{pak2013reliable}, and discrete memoryless channel (DMC) \cite{wang2016fundamental}. A positive rate has been proved to be achievable when the warden has uncertainty on his receiver noise power \cite{lee2015achieving,goeckel2016covert}, the warden does not know when the covert communication happens \cite{bash2014LPD}, or an uniformed jammer comes in to help \cite{Sobers2017Covert_J}. Most recently, \cite{BiaoHe2017on} examined the impact of noise uncertainty on covert communication by considering two practical uncertainty models. In addition, the effect of imperfect channel state information and finite blocklength on covert communication has been investigated in \cite{Shahzad2017Covert} and \cite{ShihaoYan2017Covert}, respectively, while covert communication in one-way relay networks over Rayleigh fading channels is examined in \cite{Jinsong2017Covert}.

In this work, we explore the possibilities and conditions of covert communications in quasi-static wireless fading channels, exploiting the presence of an artificial noise (AN) generated by a full-duplex receiver~\cite{fdradio_2012,fdradio_2013}. Using a full-duplex receiver offers a two-fold benefit relative to generating AN by a separate and independent jammer. Firstly, it enables a higher degree of control over the transmitted AN signals (e.g., its power), thus causing further deliberate confusion at the warden. Secondly, the cutting-edge self-interference cancellation techniques (e.g., \cite{fdradio_2012,fdradio_2013})  can be adopted to provide higher data rates for covert communications with the full-duplex receiver. Although the use of AN and jamming signals generated by the full-duplex receiver has been widely used in the literature (e.g., \cite{Zheng2013Improving,Li2012outage}) for enhancing physical-layer security, to the best of our knowledge, this use has never been studied in the context of covert communications and thus motivates our first study in this regard.

The contributions of this work are summarized as follows.
\begin{itemize}
\item We propose the use of a full-duplex receiver (Bob) to achieve covert communications in wireless fading channels, where the desired level of covertness is achieved by controlling the random transmit power of AN (from zero to a maximum value) at Bob to deliberately confuse the warden (Willie).
\item For the radiometer detector, we derive the exact expression for the detection error rate for given channel realizations from the transmitter (Alice) and Bob to Willie, based on which the optimal detection threshold is analytically determined. Surprisingly, our analysis shows that for some of these channel realizations guaranteeing a non-zero detection error rate requires the maximum transmit power of AN at Bob to approach infinity.
\item Due to the unknown wireless channels from Alice and Bob to Willie, we analytically characterize the expected detection error rate at Willie from the perspective of Alice and derive the optimal maximum transmit power of AN at Bob in order to maximize the effective covert rate for a given covertness requirement on this expected detection error rate. Interestingly, our analysis indicates that a positive effective covert rate can always be achieved for any Alice's transmit power of covert signals subject to any covertness requirement, while an upper bound on this effective covert rate exists that is achieved by increasing Alice's transmit power to infinity.
\end{itemize}

\section{System Model}
\subsection{Communication Scenario and Assumptions}
\begin{figure}[!t]
  \centering
  \includegraphics[scale=0.7]{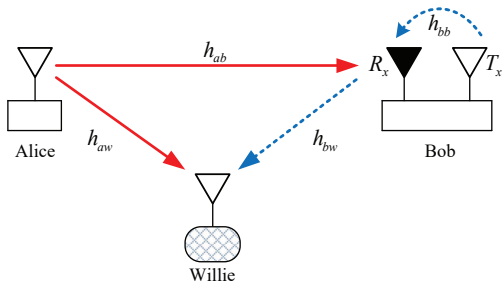}\\
  \caption{Covert communications with a full-duplex receiver.}\label{fig1}
\end{figure}

As shown in Fig.~\ref{fig1}, we consider a wireless communication scenario, where Bob (i.e., the receiver) operates in full-duplex mode and Alice (i.e., the transmitter) wants to transmit covertly to Bob with the aid of AN generated by Bob, while Willie (i.e., the warden) tries to detect this covert transmission. The channels from Alice-to-Bob, Alice-to-Willie, and Bob-to-Willie, are denoted by $h_{ab}$, $h_{aw}$, and $h_{bw}$, respectively, while the self-interference channel of Bob is denoted by $h_{bb}$. We assume that the wireless channels are subject to independent quasi-static Rayleigh fading, where the channel coefficient remains constant over one communication slot of $n$ channel uses and changes independently from slot to slot. The mean value of $|h_j|^2$ over different slots is denoted by $\lambda_j$, where the subscript $j$ can be $ab, bb, aw, bw$. Alice and Willie are assumed to have a single antenna, while besides the single receiving antenna, Bob uses an additional antenna for transmission of AN in order to deliberately confuse Willie. We assume that Bob knows $h_{ab}$, while Willie knows $h_{aw}$ and $h_{bw}$.
If Alice transmits, the signal received at Bob is given by
\begin{align} \label{y_b}
\mathbf{y}_b[i]=\sqrt{P_a}h_{ab}\mathbf{x}_a[i]+\sqrt{\phi P_b}h_{bb}\mathbf{v}_b[i]+\mathbf{n}_b[i],
\end{align}
where $\mathbf{x}_a$ is the signal transmitted by Alice satisfying $\mathbb{E}[\mathbf{x}_a[i]\mathbf{x}^{\dag}_a[i]]=1$, $i = 1, 2, \dots, n$ is the index of each channel use, $\mathbf{v}_b$ is the AN signal transmitted by Bob satisfying $\mathbb{E}[\mathbf{v}_b[i]\mathbf{v}^{\dag}_b[i]]=1$, and $\mathbf{n}_b[i]$ is the AWGN at Bob with $\sigma^2_b$ as its variance, i.e., $\mathbf{n}_b[i] \thicksim\mathcal{CN}(0,\sigma^2_b)$. Since the AN signal is known to Bob, the residual noise can be rebuilt and eliminated by self-interference cancellation. In this work, we assume that the self-interference cannot be totally cancelled and we denote the self-interference cancellation coefficient by $\phi$. Thus, $\phi=0$ refers to the ideal case, while $0<\phi\leq1$ corresponds to different self-interference cancellation levels~\cite{Everett2014Passive}. We denote the transmit powers of Alice and Bob by $P_a$ and $P_b$, respectively, where $P_a$ is fixed and publicly known by Willie and Bob. In contrast, $P_b$ changes from slot to slot and follows a continuous uniform distribution over the interval $[0, P_b^{\mathrm{max}}]$ with probability density function (pdf) given by

\begin{equation}\label{pb_pdf}
\begin{aligned}
f_{P_b}(x)=
\begin{cases}
\frac{1}{P_b^{\mathrm{max}}} & \text{if}\quad 0\leq x\leq P_b^{\mathrm{max}}, \\
0, & \text{otherwise}.
\end{cases}
\end{aligned}
\end{equation}
Since Willie possesses knowledge of $h_{aw}$ and $h_{bw}$ in the slot under consideration, for a constant transmit power at Bob, it is straightforward for him to flag a covert transmission when an additional power from Alice is received. The purpose of introducing randomness in Bob's transmit power is to create uncertainty in Willie's received power, such that Willie is unsure whether an increase in the received power is due to Alice's transmission or simply a variation in the transmit power of the Bob's AN signal. Note that we consider the uniform distribution as an initial example and other distributions will be explored in future work.

\subsection{Detection Metrics At Willie}
We focus on one communication slot, where Willie has to decide whether Alice transmitted to Bob, or not. Thus Willie faces a binary hypothesis testing problem, where the null hypothesis $\Hnull$ states that Alice did not transmit, while the alternative hypothesis $\Halt$ states that Alice did transmit to Bob. Under these hypotheses, the signal received at Willie is given by
\begin{eqnarray}\label{yw}
\mathbf{y}_w[i]\!=\!
 \left\{ \begin{aligned}
        \ &\sqrt{P_b}h_{bw}\mathbf{v}_b[i]+\mathbf{n}_w[i], ~~~~~~~~~~~~~~~~~~~~~~\Hnull, \\
        \ &\sqrt{P_a}h_{aw}\mathbf{x}_a[i]+\sqrt{P_b}h_{bw}\mathbf{v}_b[i]+\mathbf{n}_w[i],  ~~\Halt,
         \end{aligned} \right.
\end{eqnarray}
where $\mathbf{n}_w[i]$ is the AWGN at Willie with a variance of $\sigma_w^2$. We note that the value of $P_b$ in the given slot is unknown to Willie, while the value of $P_a$ is publicly known. The ultimate goal for Willie is to detect whether $\mathbf{y}_w$ comes from $\Hnull$ or $\Halt$. By application of Neyman-Pearson criterion, the optimal decision rule for Willie to minimize his detection error using the Likelihood Ratio test is given by \cite{Shahzad2017Covert}
\begin{align}\label{decisions}
T_w\mathop{\gtrless}\limits_{\Honull}^{\Hoalt}\tau,
\end{align}
where $T_w=1/n\sum_{i=1}^{n}|\mathbf{y}_w[i]|^2$ is the average power received at Willie in the slot, $\tau$ is a predetermined threshold and $\Hoalt$ and $\Honull$ are the decisions in favor of $\Halt$ and $\Hnull$ respectively. In this work, we consider an infinite number of channel uses i.e., $n \rightarrow \infty$, thus
\begin{eqnarray}\label{t_w}
T_w=
 \left\{ \begin{aligned}
        \ &P_b|h_{bw}|^2+\sigma_w^2, ~~~~~~~~~~~~~~~~~~~\Hnull, \\
        \ &P_a|h_{aw}|^2+P_b|h_{bw}|^2+\sigma_w^2,  ~~~~~\Halt.
         \end{aligned} \right.
\end{eqnarray}

Willie has to make a decision regarding Alice's action at the end of this communication slot. We define the false alarm rate (or Type I error) as the probability that Willie makes a decision in favour of $\Halt$ while $\Hnull$ is true and denote it by $\alpha= \mathcal{P}(\Hoalt|\Hnull)$. Similarly, the miss detection rate (or Type II error) is defined as the probability of Willie making a decision in favour of $\Hnull$, while $\Halt$ is true, and it is denoted by $\beta=\mathcal{P}(\Honull|\Halt)$. Under the assumption of equal \textit{a priori} probabilities of hypotheses $\Hnull$ and $\Halt$, the detection performance of Willie is measured by its detection error rate, which is defined as
\begin{align}
\xi\triangleq \alpha + \beta.
\end{align}


\section{Detection Performance at Willie}
We first analyze Willie's detection performance in terms of false alarm and miss detection rates. Next, the optimal detection threshold at Willie, minimizing the detection error rate is considered.

\subsection{False Alarm and Miss Detection Rates}
\begin{lemma}\label{lemma1}
The false alarm and miss detection rates at Willie are given by
\begin{align}
\alpha&=\left\{
  \begin{array}{ll}
    1,  &\tau<\sigma_w^2, \\
    1-\frac{\tau-\sigma_w^2}{P_b^{\mathrm{max}}|h_{bw}|^2}, & \sigma_w^2\leq\tau\leq\rho_1,\\
    0,  &\tau>\rho_1,
  \end{array}
\right. \label{PFA}\\
\beta&=\left\{
  \begin{array}{ll}
    0,  &\tau<\rho_2, \\
    \frac{\tau-\rho_2}{P_b^{\mathrm{max}}|h_{bw}|^2} , & \rho_2\leq\tau\leq \rho_3,\\
    1,  &\tau>\rho_3,
  \end{array}
\right. \label{PMD}
\end{align}
where
\begin{align}
&\rho_1\triangleq P_b^{\mathrm{max}}|h_{bw}|^2+\sigma_w^2, \notag \\
&\rho_2\triangleq P_a|h_{aw}|^2+\sigma_w^2, \notag \\
&\rho_3\triangleq P_b^{\mathrm{max}}|h_{bw}|^2+P_a|h_{aw}|^2+\sigma_w^2. \notag
\end{align}
\end{lemma}
\begin{IEEEproof}
From \eqref{t_w}, the false alarm rate is given by
\begin{align}
\alpha&=\mathcal{P}\left[P_b|h_{bw}|^2+\sigma_w^2>\tau\right] \notag \\
&=\left\{
  \begin{array}{ll}
    1,  &\tau<\sigma_w^2, \\
    \mathcal{P}\left[P_b>\frac{\tau-\sigma_w^2}{|h_{bw}|^2}\right], &\sigma_w^2\leq\tau\leq\rho_1,\\
    0,  &\tau>\rho_1.
  \end{array}
\right.
\end{align}
Similarly, the miss detection rate is given by
\begin{align}
\beta&=\mathcal{P}\left[P_a|h_{aw}|^2+P_b|h_{bw}|^2+\sigma_w^2<\tau\right] \notag \\
&=\left\{
\begin{array}{ll}
    0,  &\tau<\rho_2, \\
    \mathcal{P}\left[P_b<\frac{\tau-\rho_2}{|h_{bw}|^2}\right], &\rho_2\leq\tau\leq \rho_3,\\
    1,  &\tau>\rho_3.
\end{array}
\right.
\end{align}
where we have used the uniform pdf of $P_b$, given by $f_{P_b}(x)$ in \eqref{pb_pdf}.
\end{IEEEproof}

\subsection{Optimal Detection Threshold}
In this section, we derive the optimal value of the detection threshold $\tau$ that minimizes the detection error rate $\xi$ at Willie.

\begin{theorem}\label{theorem1}
Under the assumption of a radiometer at Willie, the optimal threshold for his detector's threshold is
\begin{eqnarray}
\tau^{\ast}\!=\!
 \left\{ \begin{aligned}
        \ &[\rho_1, \rho_2], ~~\rho_1<\rho_2, \\
        \ &[\rho_2, \rho_1],  ~~\rho_1\geq\rho_2,
         \end{aligned} \right.
\end{eqnarray}
and the corresponding detection error rate is given by
\begin{eqnarray}
\xi^{\ast}\!=\! \label{opt_xi}
 \left\{ \begin{aligned}
        \ &0, ~~~~~~~~~~~~~~~~~~~~\rho_1<\rho_2, \\
        \ &1-\frac{P_a|h_{aw}|^2}{P_b^{\mathrm{max}}|h_{bw}|^2},  ~~\rho_1\geq\rho_2,
         \end{aligned} \right.
\end{eqnarray}
where $\rho_1$ and $\rho_2$ are as defined earlier in \textit{Lemma}~\ref{lemma1}.
\end{theorem}

\begin{IEEEproof}
We first note that $\rho_3 \geq \max(\rho_1, \rho_2)$. When $\rho_1<\rho_2$, following \eqref{PFA} and \eqref{PMD}, the detection error rate at Willie is
\begin{align} \label{xi_case1}
\xi=\left\{
  \begin{array}{ll}
    1, & \tau \leq \sigma_w^2, \\
    1-\frac{\tau-\sigma_w^2}{P_b^{\mathrm{max}}|h_{bw}|^2},  &\sigma_w^2 <\tau\leq\rho_1, \\
    0,  &\rho_1 <\tau\leq\rho_2, \\
    \frac{\tau-\rho_2}{P_b^{\mathrm{max}}|h_{bw}|^2},  &\rho_2\leq\tau<\rho_3, \\
    1,  &\tau\geq\rho_3.
  \end{array}
\right.
\end{align}
Thus Willie can simply set $\tau\in\left[\rho_1, \rho_2\right]$ to ensure $\xi=0$ when $P_a|h_{aw}|^2> P_b^{\mathrm{max}}|h_{bw}|^2$, and hence can detect the covert transmission with probability one. When $\rho_1\geq\rho_2$, the detection error rate at Willie is
\begin{align} \label{xi_case2}
\xi=\left\{
  \begin{array}{ll}
    1, & \tau \leq \sigma_w^2, \\
    1-\frac{\tau-\sigma_w^2}{P_b^{\mathrm{max}}|h_{bw}|^2},  &\sigma_w^2 <\tau\leq\rho_2, \\
    1-\frac{P_a|h_{aw}|^2}{P_b^{\mathrm{max}}|h_{bw}|^2},  &\rho_2 <\tau\leq\rho_1, \\
    \frac{\tau-\rho_2}{P_b^{\mathrm{max}}|h_{bw}|^2},  &\rho_1\leq\tau<\rho_3, \\
    1,  &\tau\geq\rho_3.
  \end{array}
\right.
\end{align}
We first note that $\xi = 1$ is the worst case scenario for Willie and thus Willie does not set $\tau \leq \sigma_w^2$ or $\tau > \rho_3$. We also note that $\xi$ in \eqref{xi_case2} is a continuous decreasing function of $\tau$ when $\sigma_w^2 <\tau\leq\rho_2$. Thus, Willie will set $\rho_2$ as the threshold to minimize $\xi$ in this case. For $\rho_1\leq\tau<\rho_3$, $\xi$ is an increasing function of $\tau$. Thus Willie will set $\rho_1$ as the threshold to minimize $\xi$ if $\rho_1\leq\tau<\rho_3$. When $\rho_2\leq\tau<\rho_1$, the value of $\xi$ is constant with the minimum value $1-(P_a|h_{aw}|^2)/(P_b^{\mathrm{max}}|h_{bw}|^2)$.
\end{IEEEproof}

\begin{remark}\label{remark1}
We note here that although the noise variance at Willie i.e., $\sigma_w^2$ appears in the calculation of radiometer's optimal threshold, it does not affect the detection error rate, $\xi^{\ast}$, at Willie. On the other hand, the transmit power limit at Bob directly affects $\xi^{\ast}$, since as $P_b^{\mathrm{max}} \rightarrow \infty$, $\xi^{\ast} \rightarrow 1$.
\end{remark}

\section{Performance of Covert Communication}
In this section, we first analyze the transmission outage probability at Bob. Next, we consider optimizing the effective covert rate achieved in the system subject to a certain covert constraint.

\subsection{Transmission Outage Probability from Alice to Bob}
Following \eqref{y_b}, the signal-to-interference-plus-noise ratio (SINR) at Bob is given by
\begin{align} \label{gamma_b}
\gamma_b&=\frac{P_a|h_{ab}|^2}{\phi P_b |h_{bb}|^2+\sigma_b^2}.
\end{align}
We assume here that the transmission rate from Alice to Bob is predetermined, and is denoted by $R$. Due to the randomness in $|h_{ab}|^2$, $|h_{bb}|^2$, and $P_b$, the transmission outage probability from Alice to Bob still incurs when $C<R$, where $C$ is the channel capacity from Alice to Bob.

\begin{lemma}\label{lemma2} The transmission outage probability from Alice to Bob is given by
\begin{align} \label{outage_delta}
\delta=1-\lambda_{a\mathfrak{}b}e^{-\frac{\mu\sigma_b^2}{\lambda_{ab}}}\frac{\ln(\mu\phi \lambda_{bb}P_b^{\mathrm{max}}+\lambda_{ab})-\ln(\lambda_{ab})}{\mu\phi \lambda_{bb}P_b^{\mathrm{max}}},
\end{align}
where $ \mu=(2^{R_{ab}}-1)/P_a$.
\end{lemma}
\begin{IEEEproof} Based on the definition of the transmission outage probability, we have
\begin{align}
\delta &=\mathcal{P}\left\{\frac{P_a|h_{ab}|^2}{\phi P_b |h_{bb}|^2+\sigma_b^2}<2^{R}-1\right\} \notag \\
&=\int_0^{P_b^{\mathrm{max}}}\int_0^{+\infty}\int_0^{\mu(\phi P_b |h_{bb}|^2+\sigma_b^2)}f_{P_b}(x)f_{|h_{bb}|^2}(y)\times \notag \\
&~~~~~f_{|h_{ab}|^2}(z)dx dy dz ,
\end{align}
and solving this integral gives the desired result.
\end{IEEEproof}

\subsection{Expected Detection Error Rate}

Since Alice and Bob do not know the instantaneous realization of $h_{aw}$ or $h_{bw}$, we consider the expected value of $\xi^{\ast}$ over all realizations of $h_{aw}$ and $h_{bw}$ as the measure of covertness from Alice and Bob's perspective. We denote the expected detection error rate at Willie as $\overline{\xi^{\ast}}$. Then, the covertness requirement can be written as
 $\overline{\xi^{\ast}} \geq 1 - \epsilon$, where $\epsilon \in [0,1]$ specifies a predetermined covert constraint.

\begin{theorem}\label{theorem2}
Under the optimal threshold setting, the expected detection error rate at Willie is given by
\begin{align} \label{ave_opt_xi}
\overline{\xi^{\ast}}(t)=-t^2+t\ln t +1,
\end{align}
where
\begin{align} \label{t}
t\triangleq(P_a\lambda_{aw})/(P_a\lambda_{aw}+P_b^{\mathrm{max}}\lambda_{bw}).
\end{align}
\end{theorem}
\begin{IEEEproof}
From \eqref{opt_xi}, the expected $\xi^{\ast}$ is given by
\begin{align} \label{ave_opt_xi_before}
\overline{\xi^{\ast}}&= \mathcal{P}[\rho_1 < \rho_2] \times 0+\mathcal{P}[\rho_1\geq \rho_2]\times\mathbb{E}[\xi^{\ast}|\rho_1\geq \rho_2] \notag \\
&=\mathcal{P}[\rho_1\geq \rho_2]\times\mathbb{E}[\xi^{\ast}|\rho_1\geq \rho_2].
\end{align}
We next derive $\mathcal{P}[\rho_1\geq \rho_2]$ and $\mathbb{E}[\xi^{\ast}|\rho_1\geq \rho_2]$, which are given by
\begin{align} \label{rho1_geq_rho2}
\mathcal{P}[\rho_1\geq \rho_2]&=\mathcal{P}\left\{|h_{aw}|^2\leq \frac{P_b^{\mathrm{max}}|h_{bw}|^2}{P_a}\right\} \notag \\
&=\int_0^{+\infty}\int_0^{\frac{P_b^{\mathrm{max}}}{P_a}y}f_{|h_{bw}|^2}(x)f_{|h_{aw}|^2}(y)dx dy \notag \\
&= \frac{P_b^{\mathrm{max}}\lambda_{bw}}{P_a\lambda_{aw}+P_b^{\mathrm{max}}\lambda_{bw}},
\end{align}
and
\begin{align} \label{xi_ast_cond}
&\mathbb{E}[\xi^{\ast}|\rho_1\geq \rho_2]= 1-\frac{P_a}{P_b^{\mathrm{max}}}\mathbb{E}\left[\frac{|h_{aw}|^2}{|h_{bw}|^2}\Big{|}\rho_1\geq \rho_2\right] \notag \\
&=1-\frac{P_a}{P_b^{\mathrm{max}}}\int_0^{+\infty}\int_0^{\frac{P_b^{\mathrm{max}}}{P_a}y}\frac{x}{y}f_{|h_{aw}|^2}(x)f_{|h_{bw}|^2}(y)dx dy \notag \\
&=1-\frac{P_a\lambda_{aw}}{P_b^{\mathrm{max}}\lambda_{bw}}\Bigg\{\ln\left(1+\frac{P_b^{\mathrm{max}}\lambda_{bw}}{P_a \lambda_{aw}}\right) \notag \\
&~~~~-\frac{P_b^{\mathrm{max}}\lambda_{bw}}{P_a \lambda_{aw}+P_b^{\mathrm{max}} \lambda_{bw}}\Bigg\},
\end{align}
and substituting them into \eqref{ave_opt_xi_before} completes the proof.
\end{IEEEproof}

%
%

\subsection{Optimal AN Power and Covert Rate}
The problem of maximizing the effective covert rate achievable in the considered system subject to a certain covert constraint is given by

\begin{equation}\label{P1}
\begin{aligned}
(\mathbf{P1}) \quad \underset{P_b^{\mathrm{max}}}{\max} \quad &\overline{R}_c \\
\text{s. t.} \quad  &\overline{\xi^{\ast}} \geq 1 -\epsilon,
\end{aligned}
\end{equation}
where $\overline{R}_c \triangleq R(1-\delta)$.

\begin{theorem}\label{theorem3}
For any given transmit power $P_a$ at Alice, and a covertness constraint, $\epsilon$, the optimal choice for $P_b^{\mathrm{max}}$ to maximize the effective covert rate is given by
\begin{align} \label{P_b_max_ast}
P_b^{\mathrm{max}{\ast}}=P_a \lambda_{aw} (1-t_{\epsilon})/(t_{\epsilon}\lambda_{bw}),
\end{align}
and the maximum effective covert rate is given by
\begin{align}\label{r_c}
&\overline{R}_{c}^{\ast}=R\lambda_{a\mathfrak{}b}e^{-\frac{(2^{R}-1)\sigma_b^2}{P_a\lambda_{ab}}}\frac{\ln\left((2^{R}-1)\phi\lambda_{bb}\frac{\lambda_{aw}(1-t_{\epsilon})}{\lambda_{bw}\lambda_{ab}t_{\epsilon}}+1\right)}{(2^{R}-1)\phi\lambda_{bb}\frac{\lambda_{aw}(1-t_{\epsilon})}{\lambda_{bw}t_{\epsilon}}},
\end{align}
where $t_{\epsilon}$ is the solution of $\overline{\xi^{\ast}}(t)=1-\epsilon$ for $t$, and $\overline{\xi^{\ast}}(t)$ is as defined in \eqref{ave_opt_xi}.
\end{theorem}

\begin{IEEEproof}
We first determine the monotonicity of $\overline{\xi^{\ast}}(t)$ with respect to $t$. The first derivative is given as
\begin{align}
\frac{\partial \overline{\xi^{\ast}}(t)}{\partial t}=-2t+\ln t+1.
\end{align}
To determine the monotonicity of $\kappa(t) \triangleq -2t+\ln t+1$, we again consider the first derivative as
\begin{align}
\frac{\partial \kappa(t)}{\partial t}=-2+\frac{1}{t}.
\end{align}
Noting that $t \in [0,1)$, we conclude that $\kappa(t)$ is a monotonically increasing function with $t$ for $0\leq t< 1/2$, while it is a monotonically decreasing function with $t$ for $1/2\leq t< 1$. Since the maximum value of $\kappa(t)$ is $\kappa(1/2)=-\ln2$ for $t \in [0,1)$, we conclude that $\partial \overline{\xi^{\ast}}(t)/\partial t<0$ for $t \in [0,1)$, thus $\overline{\xi^{\ast}}(t)$ monotonically decreases with $t$. This leads to the conclusion that $\overline{\xi^{\ast}}$ monotonically increases with $P_b^{\mathrm{max}}$. Next, we will prove that the effective covert rate, $\overline{R}_c$, monotonically decreases with $P_b^{\mathrm{max}}$. Setting $\mu\phi \lambda_{bb}P_b^{\mathrm{max}}=x$, from \eqref{outage_delta}, we have $\delta = 1-\lambda_{ab}\exp[-(\mu\sigma_b^2)/\lambda_{ab}]\nu(x)$, where
\begin{align}
\nu(x) = \frac{\ln(x+\lambda_{ab})-\ln(\lambda_{ab})}{x}.
\end{align}
In order to determine the monotonicity of $\nu(x)$ with respect to $x$, we consider the first derivative as
\begin{align}
\frac{\partial \nu(x)}{\partial x}=\frac{\frac{x}{x+\lambda_{ab}}-\ln(x+\lambda_{ab})+\ln(\lambda_{ab})}{x^2},
\end{align}
where we note that whether ${\partial \nu(x)}/{\partial x} > 0$ or ${\partial \nu(x)}/{\partial x} < 0$ depends on $g(x) \triangleq x/(x+\lambda_{ab})-\ln(x+\lambda_{ab})+\ln(\lambda_{ab})$. We thus consider the first derivative of $g(x)$ as
\begin{align}
\frac{\partial g(x)}{\partial x}=-\frac{x}{(x+\lambda_{ab})^2}.
\end{align}
Noting that $x \geq 0$ and ${\partial g(x)}/{\partial x} \leq 0$, it can be seen that $g(x)$ monotonically decreases with $x$. Thus $g(x)\leq g(0)=0$ and ${\partial \nu(x)}/{\partial x} \leq 0$, which means that $\nu(x)$ is a monotonically decreasing function with $P_b^{\mathrm{max}}$, which leads to $\delta$ monotonically increasing with $P_b^{\mathrm{max}}$. As such, $\overline{R}_c$ monotonically decreases with $P_b^{\mathrm{max}}$. Since $\overline{\xi^{\ast}}$ monotonically increases with $P_b^{\mathrm{max}}$, we can obtain the optimal value of $P_b^{\mathrm{max}}$ using the constraint in \eqref{P1}. Then, using \eqref{outage_delta} and \eqref{P_b_max_ast}, we can obtain the maximum effective covert rate $\overline{R}_c^*$.
\end{IEEEproof}

\begin{corollary}\label{corollary2}
When the transmit power at Alice increases, the maximum effective covert rate approaches a fixed value given by
\begin{align}
\lim_{P_a \rightarrow \infty} \overline{R}_c^{\ast}=R\lambda_{a\mathfrak{}b}\frac{\ln\left((2^{R}-1)\phi\lambda_{bb}\frac{\lambda_{aw}(1-t_{\epsilon})}{\lambda_{bw}\lambda_{ab}t_{\epsilon}}+1\right)}{(2^{R}-1)\phi\lambda_{bb}\frac{\lambda_{aw}(1-t_{\epsilon})}{\lambda_{bw}t_{\epsilon}}},
\end{align}
\end{corollary}
\begin{IEEEproof}
Following \eqref{r_c}, $\exp\{-(2^{R}-1)\sigma_b^2/(P_a\lambda_{ab})\}\rightarrow 1$ as $P_a\rightarrow \infty$.
\end{IEEEproof}

We note that having a higher transmit power at Alice benefits Bob, but also increases Willie's probability of detecting the covert transmission. Thus to maintain the same level of covertness, Bob has to increase the AN power as well. Here, \textit{Corollary} $1$ implies that having higher transmit power at Alice has diminishing returns as the increase in self-interference due to AN's power cancels out the benefit of higher transmit power at Alice.


\section{Numerical Results and Discussion}
In this section, we present numerical results to study the performance of covert communication. For simplicity, we set $\lambda_{ab}=\lambda_{bb}=\lambda_{aw}=\lambda_{bw}=1$. As mentioned earlier, the value of $\sigma_w^2$ does not affect the detection error rate at Willie, and hence the performance of the proposed covert communication scheme. We first examine the transmission outage probability at Bob and the detection performance at Willie. Next, the impact of different system parameters on the achievable effective covert rate subject to a specific covert constraint is investigated.

\begin{figure}[t]
\centering
\includegraphics[width=3.2in, height=2.7in]{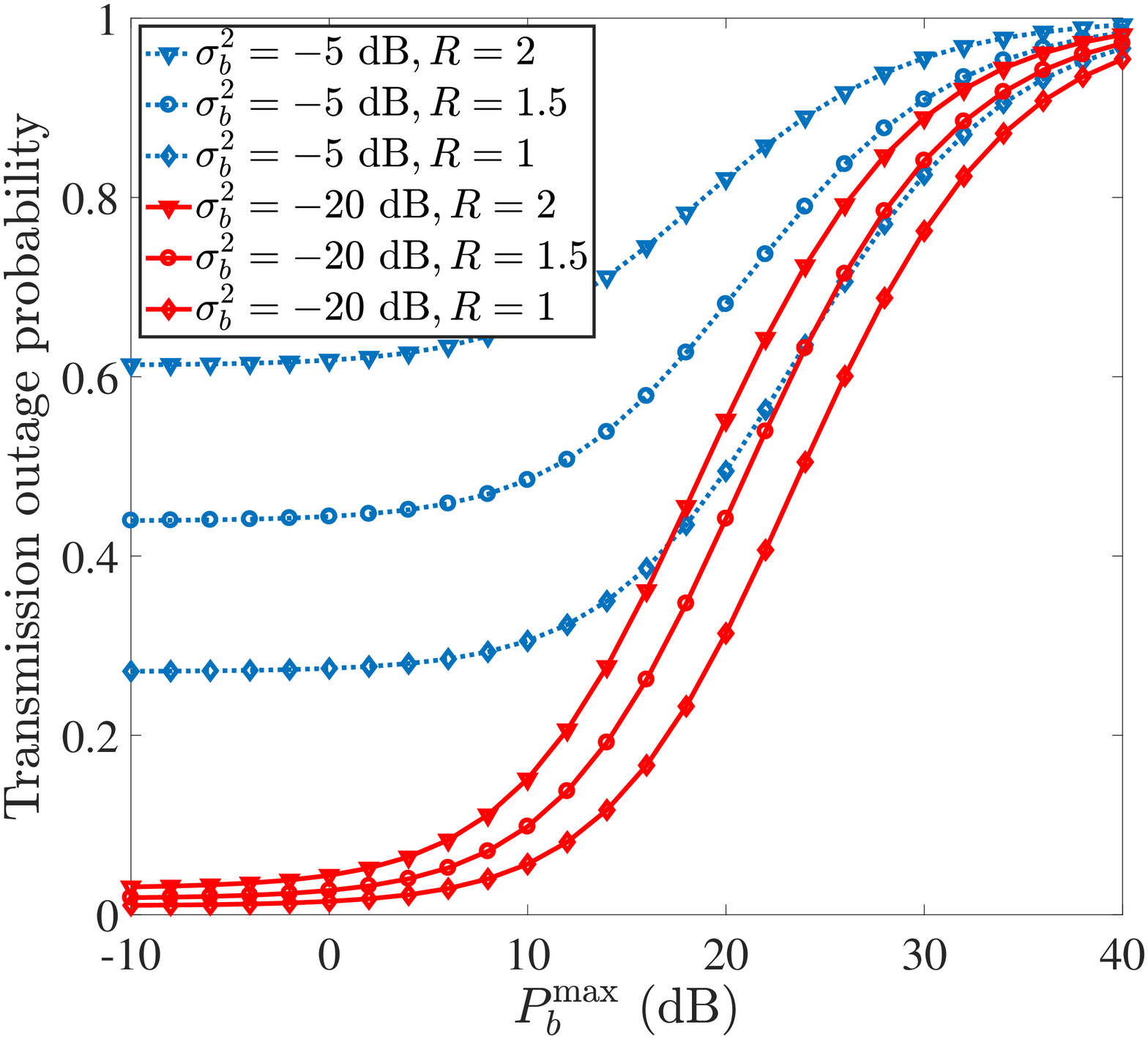}\\
\caption{Transmission outage probability $\delta$ vs. $P_b^{\mathrm{max}}$, where $P_a=0$~dB and $\phi=0.01$.}\label{fig2}
\end{figure}

In Fig.~\ref{fig2}, we illustrate the transmission outage probability $\delta$ versus $P_b^{\mathrm{max}}$ with different values of $\sigma_b^2$ and $R$. In this figure, we first observe that transmission outage probability $\delta$ is a monotonically increasing function of $P_b^{\mathrm{max}}$. This is due to the fact that the term $\phi P_b |h_{bb}|^2$, representing the self-interference, always presents a barrier for effective communication from Alice to Bob. It is also observed that $\delta$ increases with Bob's receiver noise variance $\sigma_b^2$ and the predetermined rate $R$.

\begin{figure}[t]
\centering
\includegraphics[width=3.2in, height=2.7in]{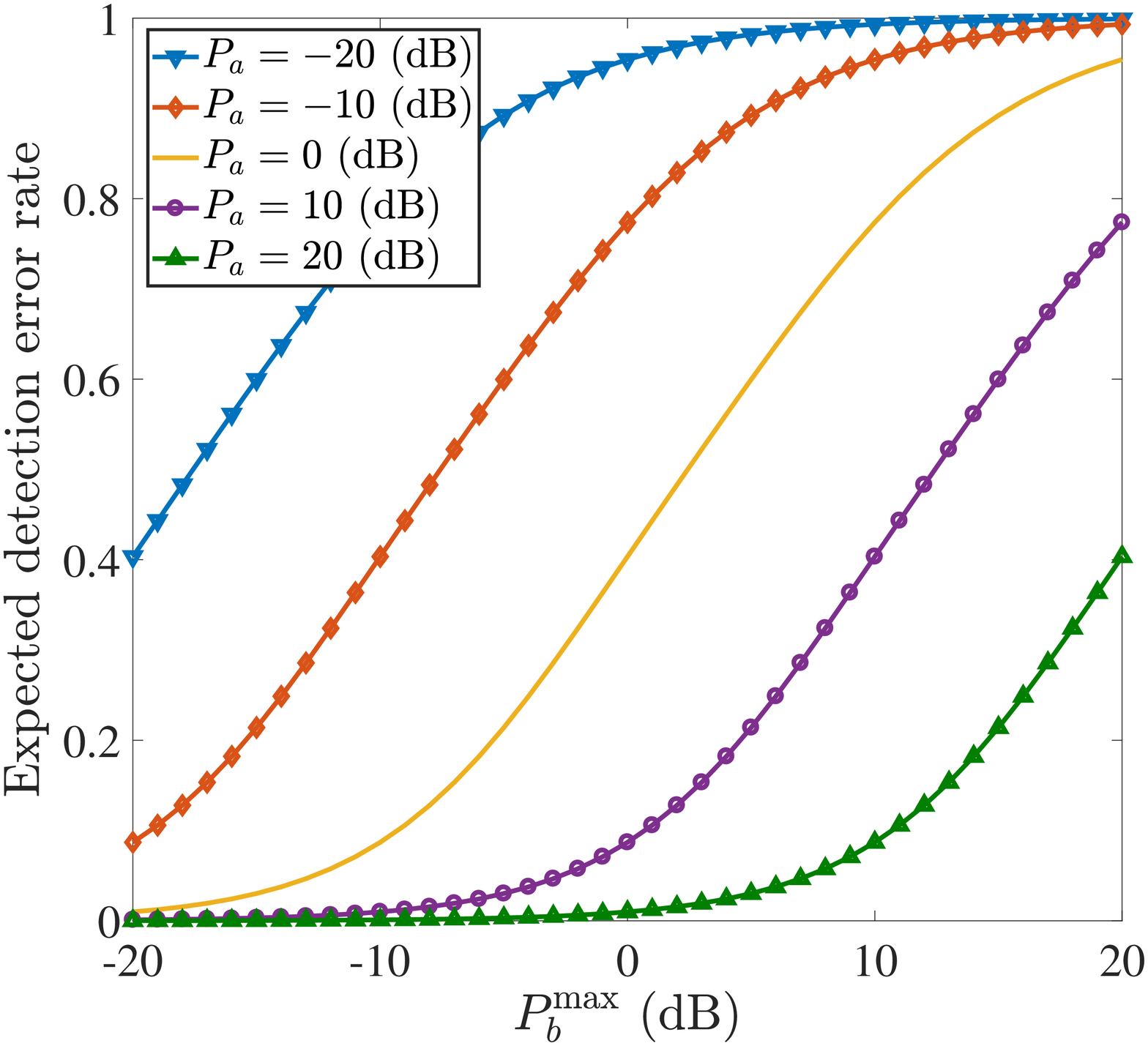}\\
\caption{Expected detection error rate at Willie, $\overline{\xi^{\ast}}$ vs. $P_b^{\mathrm{max}}$.}\label{fig3}
\end{figure}

Fig.~\ref{fig3} shows $\overline{\xi^{\ast}}$, the expected detection error rate at Willie, versus $P_b^{\mathrm{max}}$ for different values of $P_a$. In this figure, we first observe that the expected detection error rate, $\overline{\xi^{\ast}}$, is a monotonically increasing function of $P_b^{\mathrm{max}}$, as the power used to transmit the AN hinders the detection process at Willie to detect any covert transmission from Alice. It is also validated by the observation that when $P_b^{\mathrm{max}}$ is increased sufficiently, $\overline{\xi^{\ast}}\rightarrow 1$ and the covert transmission becomes harder to detect. We also observe that $\overline{\xi^{\ast}}$ is a monotonically decreasing function of $P_a$, since a higher transmit power used by Alice increases the probability of being detected by Willie.  This can also be noted from the observation that when $P_b^{\mathrm{max}}$ is sufficiently small, $\overline{\xi^{\ast}}\rightarrow 0$, and Willie can easily detect the covert transmission. These observations are also consistent with our earlier comments in \textit{Remark}~\ref{remark1}.

\begin{figure}
\centering
\includegraphics[width=3.2in, height=2.7in]{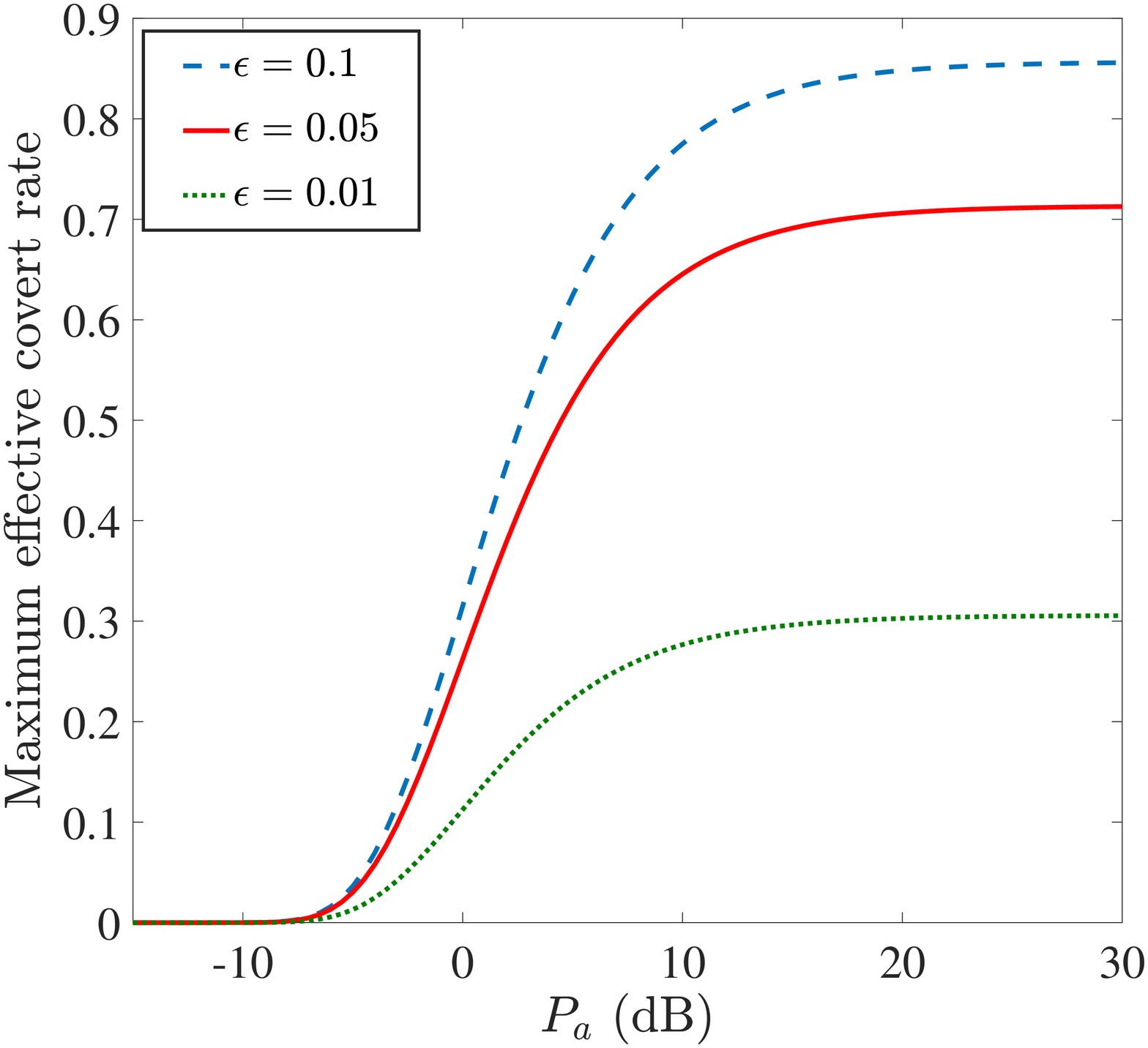}\\
\caption{Maximum effective covert rate $\overline{R}_c^*$ vs. Alice's transmit power $P_a$, where $R=1$, $\sigma_b^2=0$~dB, and $\phi=0.01$.}\label{fig4}
\end{figure}

Fig.~\ref{fig4} depicts the maximum achievable effective covert rate $\overline{R}_c^{\ast}$ versus $P_a$ for different values of $\epsilon$. We first observe that $\overline{R}_c^{\ast}$ is a monotonically increasing function of $P_a$ and $\epsilon$. This observation is consistent with our understanding of the presented system, since increasing $P_a$ increases the received SNR at Bob, while an increase in $\epsilon$ relaxes the covertness constraint.  We also observe that $\overline{R}_c^{\ast}$ approaches a fixed value when $P_a$ is sufficiently large, which validates the correctness of our \textit{Corollary}~\ref{corollary2}.

\section{Conclusion}
In this work, we analytically examined how a full-duplex receiver can aid in receiving information covertly from a transmitter by generating AN with a random transmit power. By controlling the transmit power range of the AN, a positive effective covert rate can always be achieved for any covertness requirement on the expected detection error at the warden. This work presented the first study on the possibilities and conditions of achieving covert communications by exploiting a full-duplex receiver with AN. It is expected that future work will devise improved covert transmission schemes, leading to a better tradeoff between covertness and achievable covert rates.

\section*{Acknowledgments}
This work was supported in part by the National Natural Science Foundation of China (Nos. 61472190, 61501238, and 61771244) and the Australian Research Council's Discovery Projects (DP150103905).


\end{document}